\documentclass[12pt,preprint]{aastex}

\newcommand{\ms}{\mbox{m s$^{-1}~$}}

\newcommand{\msun}{M$_{\odot}$}

\newcommand{\mjup}{M$_{\rm JUP}$}

\newcommand{\msini}{$M \sin i~$}

\lefthead{Minniti {\it et~al.\/}}
\righthead{New Magellan Planets}
\slugcomment{Accepted by The Astrophysical Journal}
\begin{document}

\title{Low Mass Companions for Five Solar-Type Stars
from the Magellan Planet Search Program\altaffilmark{1}}

\author{Dante Minniti\altaffilmark{2,6}, 
R. Paul Butler\altaffilmark{3}, 
Mercedes L\'opez-Morales\altaffilmark{3,7},
Stephen A. Shectman\altaffilmark{4}, 
Fred C. Adams\altaffilmark{5}, 
Pamela Arriagada\altaffilmark{2},
Alan P. Boss\altaffilmark{3}, 
John E. Chambers\altaffilmark{3} 
}

\authoremail{dante@astro.puc.cl}

\altaffiltext{1}{Based on observations obtained with
the Magellan Telescopes, operated by the Carnegie
Institution, Harvard University, University of Michigan,
University of Arizona, and the Massachusetts Institute
of Technology.} 

\altaffiltext{2}{Department of Astronomy, Pontificia
Universidad Cat\'olica de Chile, Casilla 306, Santiago 22, Chile}

\altaffiltext{3}{Department of Terrestrial Magnetism, Carnegie
Institution of Washington, 5241 Broad Branch Road NW,
Washington D.C. USA 20015-1305}

\altaffiltext{4}{Carnegie Observatories,
813 Santa Barbara Street, Pasadena, CA USA 91101}

\altaffiltext{5}{Astronomy Department, University of Michigan,
Ann Arbor, MI USA 48109}

\altaffiltext{6}{Specola Vaticana, V00120 Citta' del Vaticano, Italy}
\altaffiltext{7}{Hubble Fellow}

\begin{abstract}

We report low mass companions orbiting five Solar-type stars
that have emerged from the Magellan precision Doppler velocity
survey, with minimum (\msini) masses ranging from 1.2 to 25 \mjup.
These nearby target stars range from mildly metal-poor to metal-rich, and
appear to have low chromospheric activity.
The companions to the brightest two of these stars have previously been 
reported from the CORALIE survey. Four of these companions
(HD 48265-b, HD 143361-b, HD 28185-b, HD 111232-b) are low-mass
Jupiter-like planets in eccentric intermediate and long-period 
orbits. On the other hand, the companion to HD 43848 appears to be a 
long period brown dwarf in a very eccentric orbit.

\end{abstract}

\keywords{planetary systems -- stars: individual (HD 48265,
HD 143361, HD 43848, HD 28185, HD 111232)}

\section{Introduction}
\label{intro}

A decade after the milestone discoveries of the first
extrasolar planet (Mayor \& Queloz 1995), the first
planetary system (Butler et al. 1999), and the first
transiting planet (Henry et al. 2000; Charbonneau et al. 2000)
precision Doppler surveys remain the dominant method for
finding extrasolar planets orbiting nearby stars.
Most of the planets orbiting stars within 100pc have been found
from precision velocity surveys.  Current surveys now cover nearly
all main sequence G and K stars within 50 pc.  Northern hemisphere
planet radial velocity 
surveys include the Lick 3-m, Elodie 1.93-m, Keck 10-m,
and the HET 11-m.  Surveys in the southern hemisphere include the
HARPS 3.6-m, AAT 3.9-m, and the Coralie 1.2-m. 

With the unexpected diversity of extrasolar planets found to date
(Butler et al. 2006), every new planet continues to contribute
to our understanding.  Planets around the stars closer than 100 pc are
especially valuable as these are the stars that will be prime
targets for studies
by future interferometry and direct imaging missions.

In this work we present the detection of low-mass companions to
five nearby solar-type stars, including four of planetary mass,
and one likely brown dwarf.  The brightest two of the objects 
reported here have
previously been detected by the CORALIE team (Santos et al. 2001;
Mayor et al. 2004).  These sub-stellar companions were
found by the Magellan precision Doppler survey, which has been operated
over the past 4 years, and is focused on late F, G, K, and M dwarfs
within 50 pc.  This program is complementary to the Magellan N2K search
for "Hot Jupiters" (Fischer et al. 2003; Lopez-Morales et al.  2008),
which surveys metal rich stars primarily beyond 50 pc, with time
allocated by the Chilean National TAC.
The same hardware and software reduction packages
are used by both programs.

This paper is organized as follows.
Section 2 provides an overview of the Magellan Planet Search Program.
Section 3 contains a summary of properties of the 5 stars, Doppler
velocity measurements, and best-fit orbital solutions for companions.
A discussion is given in Section 4.

\section{The Magellan Planet Search Program}
\label{obs_01}

The Magellan Planet Search Program makes use of the MIKE echelle
spectrograph (Bernstein et al. 2003) on the 6.5-m Magellan II (Clay)
telescope.  The resolution of these spectra is R $\sim$ 50000,
spanning wavelengths from 3900--6200 \AA, divided into a blue and a red
CCD.  Wavelength calibration
is carried out by means of an iodine absorption cell
(Marcy \& Butler 1992) which superimposes a reference iodine
spectrum directly on the stellar spectra. ThAr lamps are not used in any part of the reduction.  The wavelength solution is derived directly from the Iodine spectrum, using the spectral ``atlas" of this cell
taken with the NIST FTS spectrometer at a resolution of 1,000,000 and a S/N
of 1,000 as the metric (Butler et al. 1996; Valenti et al. 1995). The standard Iodine technique works between 5000 \AA~and 6200 \AA.  Below 5000 \AA,
Iodine molecules disassociate, so there are no lines.  Above 6200 \AA~ Iodine lines are
too weak to be useful, unless the Iodine cell is heated to several
hundred degrees C. Thus, only the spectral orders of  the 
MIKE red CCD are used to measure the radial velocities, while
the blue orders are used to monitor the activity of the CaII H and K lines.

We have monitored a number of stable main
sequence stars of spectral types ranging from late F to mid K.
Several example of these RV stable stars are shown in Figures 1 and 2. 
This system currently achieves
measurement precision of 5 \ms, as demonstrated by these figures.

The Magellan Precision Doppler Survey began in December 2004, and
is currently surveying about 400 main sequence dwarfs ranging in
spectral type from late F to mid--M.  The spectrum of stars earlier
than F7 do not contain enough Doppler information to achieve precision
of 5 \ms, while stars later than M5 are too faint even for a 6.5-m
telescope. Magellan/MIKE can reach 5 m/s for stars as faint at V=9 with 10 minute
exposures..  The stars in the Magellan program have been chosen to minimize
overlap with the AAT 3.9-m and Keck 10-m surveys.  
Subgiants have not been removed.  Stellar jitter for
subgiants is small, $\lesssim$ 5 m/s.  Stars more than 2 magnitudes
above the main sequence have much larger jitter, thus
have been removed from the observing list based on Hipparcos
distances (Perryman et al. 1997, ESA 1997).

Stars with known stellar companions within 2 arcsec are also
removed from the observing list as it is operationally difficult
to get an uncontaminated spectrum of a star with a nearby
companion.  Otherwise there is no bias against observing
multiple stars.  The Magellan target stars also contain no bias
against brown dwarf companions.

\section{Companions from the Magellan Survey}
\label{obs_02}

This paper reports on four planet--mass candidates and a brown 
dwarf candidate that have
emerged from the Magellan survey.  The stellar properties of
the five host stars are given in Table 1.  The first two columns
provide the HD catalog number and  the Hipparcos catalog number
respectively.  Spectral types are from a calibration of $B-V$
and Hipparcos derived absolute magnitudes.
The stellar masses are estimated by interpolation of evolutionary
tracks (Fuhrmann 1998, Fuhrmann et al. 1997).  The [Fe/H] values
listed in Table 1 are drawn from a variety of sources (given below).

Figure 3 shows the K line for the 5 stars reported in this paper, 
in ascending order of B-V.  The Sun (bottom) is shown for
comparison.  Mt. Wilson S values (Baliunas et al. 1995) have
been measured from the H \& K lines in the Magellan spectra.
The log(R'$_{\rm HK}$) values are listed in Table 1.
These stars are all chromospherically quiet. Therefore, we can
constrain the expected photosperic radial velocity jitter ($< 3$ m/s), 
and reject activity as the cause of the observed
radial velocity variations of these stars discussed below.

The orbital parameters were obtained by fitting single planet Keplerian orbits to the radial-velocity data (Lopez-Morales et al. 2008).
The orbital parameters of the five companions are listed
in Table 2, while the individual Magellan Doppler velocity
measurements are listed in Tables 3 through 7.  The properties of
the host stars and of their companions are discussed in turn below.

\subsection{HD 111232}

HD 111232 is a G5V dwarf with $V=7.59$ and $B-V=0.701$. 
The \emph{Hipparcos} parallax (Perryman et al. 1997)
gives a distance of  29.41 pc and an absolute visual magnitude
$M_V=5.28$. The star is chromospherically
quiet, with log(R'$_{\rm HK}$)$=-$5.13. 
Its metallicity is [Fe/H]$=-0.36$ (Nordstrom et al. 2004).

Based on 38 observations, the CORALIE team (Mayor et al. 2004)
discovered a planet orbiting HD 111232 with an orbital period
$P$ $=$ $1143$d, a radial velocity semi-amplitude of 
$K$ $=$ $159$ /ms, and an
eccentricity $e$ $=$ 0.20.  The RMS to the best-fit Keplerian
orbit is 7.5 \ms.

Fifteen Magellan Doppler velocity observations of HD 111232
spanning 3.6 years have been made, as shown in Figure 4
and listed in Table 3.  The observations span a full orbital
period.  The period of the best-fit Keplerian orbit
is $P=1118\pm 30$ days,
the semi-amplitude is $K=162$ \ms, and the eccentricity
is $e=0.19\pm 0.10$, in agreement with the CORALIE results. The RMS
of the velocity residuals to the Keplerian fit is 4.70 \ms.
The reduced $\chi_{\nu}$ of the Keplerian fit is 1.48.
Given the mass of the star $\rm{M_{\odot}=1.06}$, the minimum
mass of the companion is \msini=8.1 M$_{Jup}$, and the orbital
semi-major axis is 2.1 AU.

\subsection{HD 48265}

HD 48265 is a G5 IV/V star with $V=8.05$ and $B-V=0.747$. The \emph{Hipparcos}
parallax (Perryman et al. 1997) gives a distance of  87.4 pc and
an absolute visual magnitude, $M_V=3.34$.  
HD 48265 is chromospherically quiet, with log(R'$_{\rm HK}$)=-4.93.
Its metallicity is [Fe/H]$=0.17$ (Nordstrom et al. 2004).

Seventeen Magellan Doppler velocity observations have been
made of HD 48265 spanning 4.4 years, as shown in Figure 5
and listed in Table 4.  The observations span 2 orbital
periods.  The semi-amplitude of the best-fit Keplerian orbit
is $K=29$ \ms, the period is $P=762$ days and the eccentricity is $e=0.24$. 
The RMS of the velocity residuals to the Keplerian orbital fit is 5.14 \ms.
The reduced $\chi_{\nu}$ of the Keplerian orbital fit is 1.37.
Assuming a stellar mass of M=0.93 \msun we derive a minimum
mass of \msini=1.3 $M_{Jup}$ and an orbital semi-major axis of 1.3 AU.

\subsection{HD 28185}
HD 28185 is a G0 V star with $V=7.80$ and $B-V=0.750$. The \emph{Hipparcos}
parallax (Perryman et al. 1997) gives a distance of  39.6 pc and
an absolute visual magnitude, $M_V=4.81$.
HD 28185 is chromospherically quiet with log(R'$_{\rm HK}$)=-4.81.
Its metallicity is [Fe/H]$=0.24$ (Fischer et al. 2005).

The planet orbiting HD 28185 was discovered by the CORALIE extra-solar
planet search (Santos et al. 2001), using 40 high-precision
radial velocity measurements over a 2-year period. 
The best-fit Keplerian solution to the CORALIE data has an
RMS of 10 \ms, and yields a period ($P$) of 383 days, a semi-amplitude
($K$) of 161 \ms, and an eccentricity $e=0.07$.
They marginally detect a long term linear drift suggesting the presence
of a second long period companion.

Fifteen Magellan Doppler velocity observations have been
made of HD 28185 over 3.6 years, as shown in Figure 6 and listed
in Table 5.  These observations span three orbital periods.  
The best-fit Keplerian orbit to the Magellan data yields a period
$P = 379 \pm 2$ days, a semi-amplitude ($K$) of 163 \ms, and an eccentricity
$e = 0.05\pm 0.03$, in agreement with the CORALIE results.
The RMS of the velocity residuals to the Keplerian orbital fit is 7.33 \ms.
The reduced $\chi_{\nu}$ of the Keplerian orbital fit is 1.51.
Given the stellar mass $\rm{M_{\odot}=1.24}$, the minimum mass
of the planet is \msini=6.7 $M_{Jup}$ with an orbital semi-major
axis of 1.1 AU.

The best-fit Keplerian orbit to the Magellan data does not include a linear
trend.  The largest term trend consistent with the Magellan data is
5 \ms per year.  More observations will be needed to constraint the
possible long term trend reported by the CORALIE team.

\subsection{HD 143361}

HD 143361 is a G0 V star with $V=9.20$ and $B-V=0.773$. The \emph{Hipparcos}
parallax (Perryman et al. 1997) gives a distance of 59.35 pc and an
absolute visual magnitude, $M_V=5.33$.  HD 143361 is chromospherically quiescent
with log(R'$_{\rm HK}$)=-4.92.
Its metallicity is [Fe/H]$=0.29$ (Nordstrom et al. 2004).
We note that this star has similar $B-V$ color and metallicity as 
HD 28185, but that they differ more than
20\% in mass and about 0.5 mag in absolute luminosity.

In our survey, twelve Magellan Doppler velocity observations have been
made of HD 143361 over 4.49 years, as shown in Figure 7
and listed in Table 6.  The observations span more than one orbital
period.  The semi-amplitude of the best-fit Keplerian orbit is 
$K=63$ \ms,
the orbital period is $P=1086$ days and the eccentricity is $e=0.18$.
The RMS of the velocity residuals to the Keplerian orbital
fit is 3.88 \ms. The reduced $\chi_{\nu}^2$ of the Keplerian orbital
fit is 1.18. 
Given the stellar mass $\rm{M}=1.00$\msun, the derived minimum
mass of the companion is \msini=3.1 $M_{Jup}$ and the
semimajor axis is 2.1 AU.

\subsection{HD 43848}
	
HD 43848 is a G2 V star with $V=8.65$ and $B-V=0.927$. The \emph{Hipparcos}
(Perryman et al. 1997) derived distance is 37.05 pc and the absolute
visual magnitude is $M_V=5.80$.  The star is chromospherically quiet, 
with log(R'$_{\rm HK}$)=-4.97. The reemission seen in Figure 3 
is negligible, and consistent with the low activity index for this 
star, and it has no impact on the radial velocity signal interpretation.
Its metallicity is [Fe/H]$=-0.03$ (Prieto \& Lambert 1999).

Ten Magellan Doppler velocity observations have been
made of HD 43848  over 4.1 years, as shown in Figure 8
and listed in Table 7.  The semi-amplitude of the best-fit 
Keplerian orbit 
is $K=544$ \ms, the period is $P=2371$ days and the eccentricity is $e=0.69$.
The RMS of the velocity residuals to the Keplerian orbital
fit is 7.16 \ms.  The reduced $\chi_{\nu}$ of the Keplerian orbital
fit is 1.77. 
Given the stellar mass $\rm{M_{\odot}=0.93}$, the derived minimum mass
is \msini=25 $M_{Jup}$ with an orbital semi-major
axis of 3.4 AU.
This is likely a brown dwarf, unless the orbital inclination is
smaller than about 18 degrees, in which case it could be a low mass
M-dwarf. Nevertheless, such inclination is improbable (less than 5\%).

A low mass companion of spectral type M3.5--M6.5 for this star was previously
discovered by Eggenberger et al. (2007) using astrometric measurements with
VLT/NACO. Their estimated companion's mass gives $\rm{M=0.14 M_{\odot}}$
with a separation of 30.9 AU.  The time span of the Magellan observations
is too short to detect the presence of this long period M dwarf.   

Because of its high eccentricity ($e=0.69$), the companion for
HD 43848 would have a maximum separation from the primary of nearly 6 AU at
apastron. At a distance of 37 pc, this yields an angular separation of 
0.16 arcsec. We estimate a magnitude of $K\sim 6.6$ for the primary, 
but the magnitude of the brown dwarf companion 
is more uncertain. A dusty brown dwarf would be too faint, while for
a clear brown dwarf we expect an absolute magnitude of $M_K\sim 10$,
and apparent magnitude $K\sim 13$. 
The detection of this companion in the near-IR seems to be at the limit
using current instrumentation at HST or VLT. 
However, this would be a prime target for detection
with new instrumentation such as PRIMA at the VLTI.

We have also checked the dispersion in the Hipparcos measurements
for this star in comparison with stars of similar magnitudes located
at similar distances. However, the errors are comparable, and there
is no evidence for higher dispersion that would indicate the presence of
a massive companion, and at the same time the data are not
good enough to secure the substellar nature of the companion.

\section{Discussion}

A new radial velocity search for extrasolar planets
carried out at the Magellan Clay telescope with MIKE is presented.
We demonstrate long term precision of 5 m/s (Figures 1 and 2).

This paper also reports the detection of four planet mass companions and
one prospective brown dwarf candidate orbiting nearby G dwarfs.  Two
of the planets belonging to the brightest two candidates of the present
sample were initially discovered by the CORALIE team.  These
stars range from mildly metal poor to metal rich.  

Figure 9 shown the eccentricity $versus$ period for the known
planets, highlighting the five objects studied here. The planet mass
companions have circular to mildly eccentric orbits with periods
ranging from 1 to 3 years. Note the extreme location of the
candidate brown dwarf HD 43848-b, with high mass and high eccentricity.
This figure suggests that objects like this might not be uncommon at 
large periods. Also, this object would be a prime target for 
detection with the future interferometric facilities.

Figure 10 shows the position of known extrasolar planetary system in the
star mass $vs$ apparent magnitude diagram. The new planets from our survey   
(this work plus Lopez-Moralez et al. 2008) are shown with large full
circles. These first Magellan discoveries are among the faintest targets
surveyed for planets, probing deeper in the Solar Neighborhood.
Note also the group of low mass stars, a range that is 
also accessible by Magellan.
We expect that the Magellan telescopes will keep on contributing to
the discovery of southern extrasolar planets.

The Magellan Planet Search Program has designed and built a custom
Planet Finder Spectrograph (Crane et al. 2006; Crane et al. 2008)
for the Clay Telescope.  This new instrument will be installed
at the end of 2008.  Advantages of this new spectrograph include
higher throughput, higher resolution, active and passive temperature
stabilization, fixed format, and all optics optimized for the
Iodine region (5000 to 6200 Angstroms).  The goal is to reach
precision of 1 \ms.  

The Magellan Planet Search Program is also collaborating with
Greg Henry and the Tennessee State University Automated Astronomy
Group to install two dedicated 0.8-m robotic photometry telescopes
at Las Campanas to monitor nearby southern hemisphere planet
search stars with a precision of 1 milli-mag.  We expect these
telescopes will be operational in early 2009.

\acknowledgements
We are thankful to Dr. Andrea Richichi (ESO),
and to Dr. John Debes (DTM) for useful comments
about the possible detection of HD 43848-b.
We are grateful to the NIST atomic spectroscopy staff,
in particular to Dr. Gillian Nave and Dr. Craig Sansonetti, for
their expert assistance in calibrating our Iodine cell with the
NIST FTS
RPB gratefully acknowledges support from NASA OSS grant NNX07AR4OG.
MLM acknowledges support provided by NASA through Hubble Fellowship
grant HF-01210.01-A awarded by the STScI, which is operated by the
AURA, Inc., for NASA, under contract NAS5-26555.  
DM and PA are supported by the Basal CATA PFB 06/09, and 
FONDAP Center for Astrophysics 15010003.
This paper has made use of the Simbad and NASA ADS data bases.

\clearpage

\begin{deluxetable}{rrrllllll}
\tablecaption{Stellar Properties}
\label{candid}
\tablewidth{0pt}
\tablehead{
\colhead{Star} & \colhead{Star} & \colhead{Spec} & \colhead{M$_{\rm Star}$} & \colhead{V} & \colhead{B-V} & \colhead{log(R'$_{\rm HK}$)} & \colhead{[Fe/H]} & \colhead{d} \\
\colhead{(HD)} & \colhead{(Hipp)} &\colhead{type} & \colhead{(M$_{\odot}$)} & \colhead{(mag)}  &  &  &  & (pc) 
} 
\startdata
  111232 &   62534 & G5 ~V & 0.75 &7.59 & 0.701 & -5.13 & -0.36 & 28.9 \\
  48265 &   31895 & G5 ~V & 0.93 & 8.05 & 0.747 & -4.93 & 0.17 & 87.41 \\
 28185 &  20723 & G0 ~V & 1.24 & 7.80 & 0.750 & -4.81 & 0.24 & 39.56 \\
 143361 &  78521 & G0 ~V & 0.95 & 9.20 & 0.773 & -4.92 & 0.29 & 59.35 \\
 43848 &  29804 & G2 ~V & 0.89 & 8.65 & 0.927 & -4.97 & -0.03 & 37.05 \\
\enddata
\end{deluxetable}

\clearpage

\begin{deluxetable}{rlllllllll}
\rotate
\tablecaption{Orbital Parameters}
\label{candid}
\tablewidth{0pt}
\tablehead{
\colhead{Star}  & \colhead{Period} & \colhead{$K$} & \colhead{$e$} & \colhead{$\omega$} & \colhead{$T_0$} & \colhead{\msini} & \colhead{$a$} & \colhead{N$_{obs}$} & \colhead{RMS} 
\\
\colhead{(HD)} & \colhead{(days)} & \colhead{(\ms)} &\colhead{ } & \colhead{(degrees)} & \colhead{(JD-2450000)}  & \colhead{(\mjup)} & {(AU)} & \colhead{ } & \colhead{(\ms)}
} 
\startdata
111232  &  1118 (30)  & 162 (20)  & 0.19 (0.10) & 82 (15) & 3501 (35) & 6.7 & 1.9 & 15 & 4.70 \\
48265 &  762 (50)  & 29   (6)  & 0.24 (0.10) &  289 (50) & 2892 (100)  & 1.2  & 1.6 & 17 & 5.14 \\
28185 &  379.0 (2)  & 163.5 (3)  & 0.05 (0.03) & 44 (2) & 4230.5 (2) & 6.7 & 1.1 & 15 & 7.33 \\
143361 & 1086 (90) & 63 (20) & 0.18 (0.10) & 187 (32) & 3439 (190) & 3.0 & 2.0 & 12 & 3.88 \\
43848&  2371  (840) & 544  (200)  & 0.69 (0.12) & 229 (9) & 3227 (65) & 25 & 3.4 & 10 & 7.16 \\
\enddata
\end{deluxetable}

\clearpage

\begin{deluxetable}{rrr}
\tablecaption{Velocities for HD 111232}
\label{vel111232}
\tablewidth{0pt}
\tablehead{
JD & RV & error \\
(-2450000)   &  (m s$^{-1}$) & (m s$^{-1}$)
}
\startdata
   863.4875  &   -28.5  &  7.5 \\
  1042.7591  &    90.0  &  7.8 \\
  1127.7398  &   123.6  &  3.3 \\
  1455.8151  &   108.0  &  3.2 \\
  1542.5330  &    -6.8  &  3.5 \\
  1756.8718  &  -128.6  &  2.8 \\
  1757.8713  &  -129.6  &  2.7 \\
  1774.8630  &  -137.6  &  3.3 \\
  1775.8698  &  -130.2  &  3.7 \\
  1783.8728  &  -128.7  &  2.8 \\
  1785.8745  &  -125.2  &  2.8 \\
  1811.8014  &  -116.2  &  2.9 \\
  1872.6422  &  -101.0  &  2.9 \\
  1896.5819  &   -80.0  &  3.5 \\
  2189.8023  &    89.6  &  3.0 \\
\enddata
\end{deluxetable}

\clearpage

\begin{deluxetable}{rrr}
\tablecaption{Velocities for HD 48265}
\label{vel48265}
\tablewidth{0pt}
\tablehead{
JD & RV & error \\
(-2450000)   &  (m s$^{-1}$) & (m s$^{-1}$) 
}
\startdata
   920.8629  &    20.6  &  6.9 \\
  1431.6073  &   -30.2  &  2.5 \\
  1455.5573  &   -35.0  &  2.6 \\
  1685.8138  &    15.0  &  2.9 \\
  1774.6744  &    25.0  &  3.3 \\
  1775.6763  &    12.4  &  5.2 \\
  1784.6887  &    25.4  &  2.6 \\
  1811.5943  &    24.3  &  2.5 \\
  1987.9168  &    -7.7  &  2.9 \\
  2078.7826  &   -17.9  &  3.0 \\
  2081.7133  &   -23.4  &  2.8 \\
  2137.6436  &   -20.7  &  2.5 \\
  2138.6670  &   -28.7  &  2.6 \\
  2189.5669  &   -29.5  &  4.4 \\
  2483.6148  &    28.5  &  3.0 \\
  2501.6725  &    21.5  &  2.4 \\
  2522.6219  &    25.4  &  2.7 \\
\enddata
\end{deluxetable}

\clearpage

\begin{deluxetable}{rrr}
\tablecaption{Velocities for HD 28185}
\label{vel28185}
\tablewidth{0pt}
\tablehead{
JD & RV & error \\
(-2450000)   &  (m s$^{-1}$) & (m s$^{-1}$) 
}
\startdata
   627.6637  &   120.0  &  5.2 \\
   663.5918  &   158.0  &  6.3 \\
   918.7297  &   -92.5  &  4.9 \\
   920.7447  &   -87.8  &  4.7 \\
   981.7281  &    72.2  &  5.5 \\
  1306.8369  &   -54.6  &  5.5 \\
  1307.8161  &   -72.9  &  5.5 \\
  1431.5270  &   167.8  &  3.9 \\
  1455.4900  &   147.1  &  4.1 \\
  1656.8504  &  -123.6  &  4.2 \\
  1657.8599  &  -121.7  &  4.2 \\
  1685.7325  &   -60.5  &  4.1 \\
  1757.6165  &   104.4  &  4.1 \\
  2077.6809  &   -34.2  &  3.9 \\
  2079.7824  &   -40.0  &  3.8 \\
\enddata
\end{deluxetable}

\clearpage

\begin{deluxetable}{rrrr}
\tablecaption{Velocities for HD 143361}
\label{vel143361}
\tablewidth{0pt}
\tablehead{
JD & RV & error \\
(-2450000)   &  (m s$^{-1}$) & (m s$^{-1}$) 
}
\startdata
   864.5793  &    62.1  &  6.8 \\
  1130.8371  &    19.5  &  2.4 \\
  1872.7370  &    53.2  &  2.5 \\
  1987.5051  &    62.2  &  3.0 \\
  1988.4946  &    51.3  &  2.6 \\
  2190.8055  &    25.7  &  3.2 \\
  2217.8473  &    20.0  &  3.0 \\
  2277.6582  &    -0.4  &  3.2 \\
  2299.5595  &    -8.3  &  2.6 \\
  2300.5804  &    -7.7  &  2.2 \\
  2339.5005  &   -23.6  &  3.2 \\
  2501.8720  &   -68.8  &  2.9 \\
\enddata
\end{deluxetable}

\clearpage

\begin{deluxetable}{rrrr}
\tablecaption{Velocities for HD 43848}
\label{vel43848}
\tablewidth{0pt}
\tablehead{
JD & RV & error \\
(-2450000)   &  (m s$^{-1}$) & (m s$^{-1}$) 
}
\startdata
   626.7557  &   124.1  &  5.8 \\
   920.8278  &   -90.8  &  6.0 \\
   978.7597  &  -173.4  &  7.2 \\
  1431.5773  &   498.7  &  3.5 \\
  1455.5289  &   521.2  &  3.7 \\
  1774.6672  &   479.7  &  4.2 \\
  1775.6690  &   486.9  &  7.1 \\
  1783.6628  &   483.0  &  3.6 \\
  1810.5720  &   486.9  &  3.5 \\
  2136.6677  &   406.5  &  3.7 \\
\enddata
\end{deluxetable}

\clearpage
\clearpage
\begin{figure}
\epsscale{.8}
\plotone{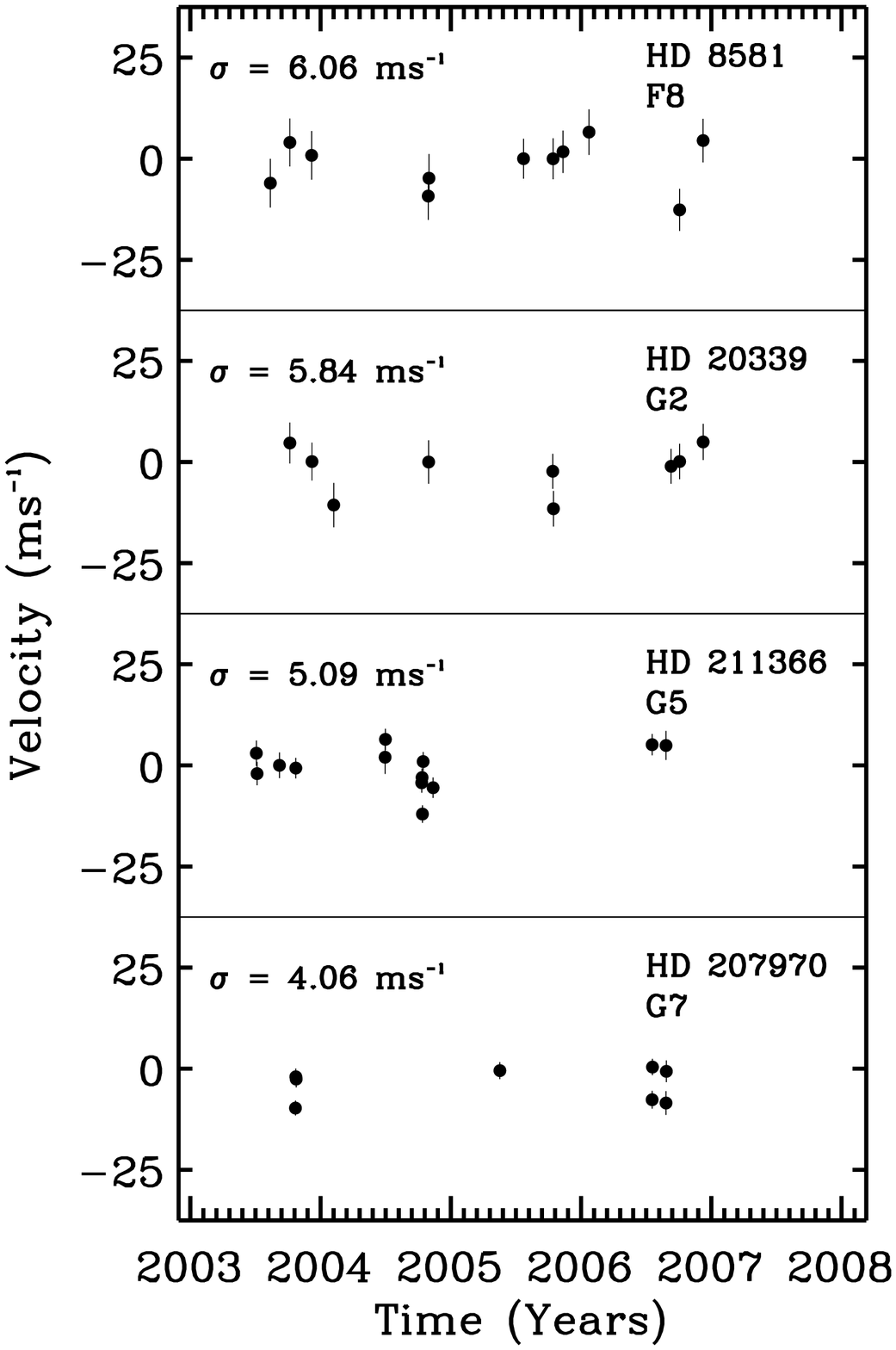}
\caption{Four MIKE stable stars with spectral types
ranging from late F to mid G.}
\label{stable1}
\end{figure}

\begin{figure}
\plotone{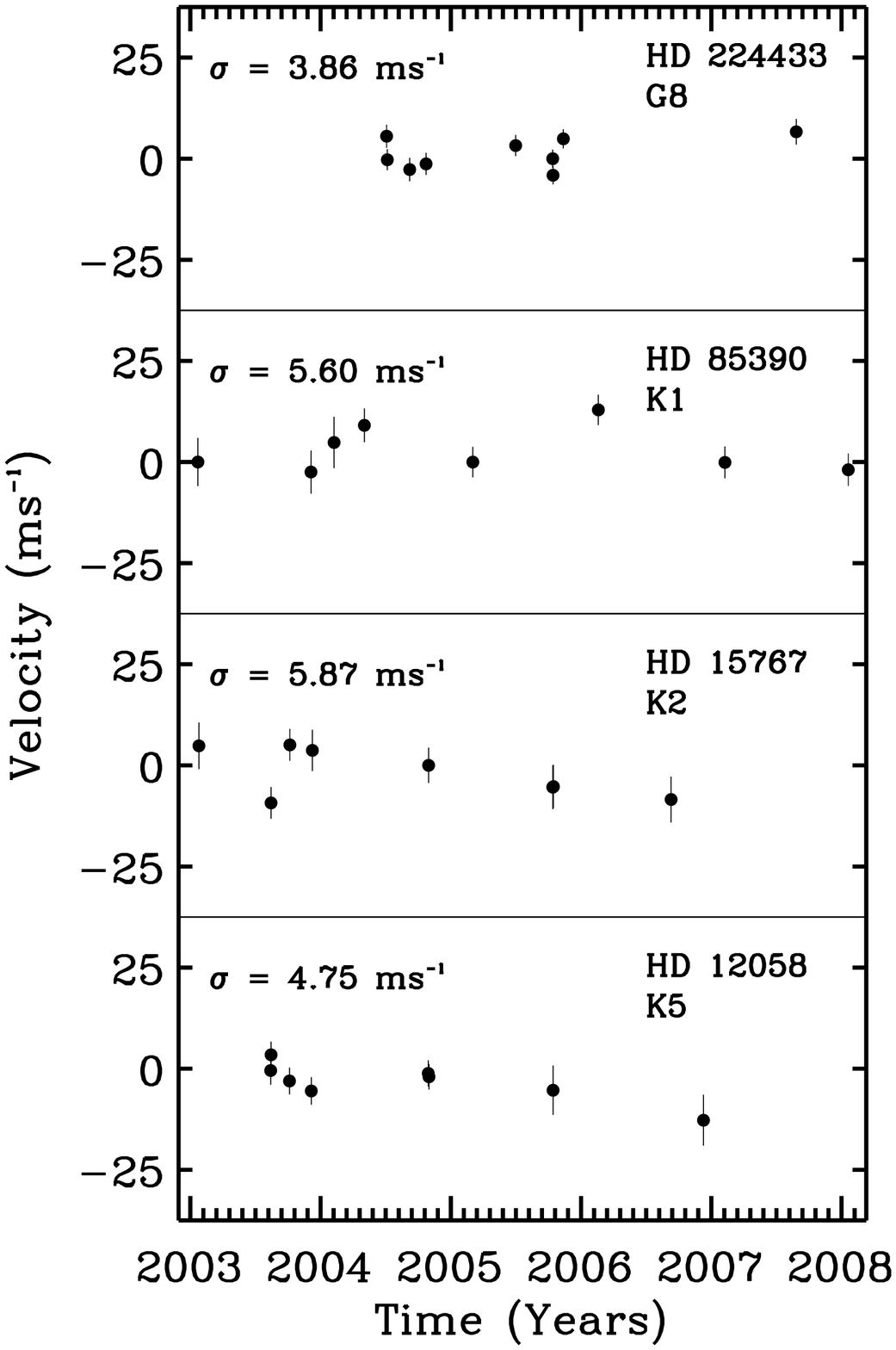}
\caption{Four MIKE stable stars with spectral types
ranging from late G to mid K.}
\label{stable2}
\end{figure}

\clearpage
\begin{figure}
\includegraphics[angle=90,width=\textwidth]{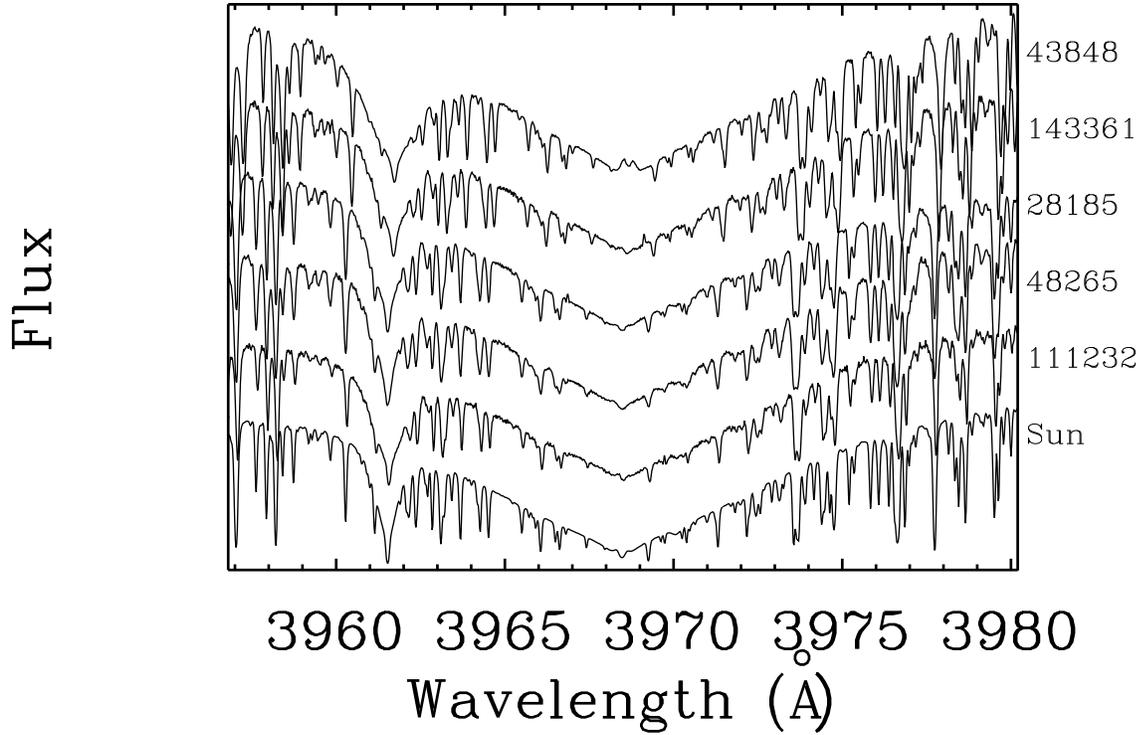}
\caption{Ca II H line cores for the five target G dwarfs in ascending
order of $B-V$.  The HD catalog number of each star is
shown along the right edge.  The Sun is shown for comparison.
The R'$_{\rm HK}$ values 
derived from the H\&K lines are similar to the Sun, indicating
rotation periods of 25 d or longer and photospheric Doppler
``jitter'' of 3 \ms or less.  
}
\label{H_Gdwarfs}
\end{figure}

\begin{figure}
\includegraphics[angle=90,width=\textwidth]{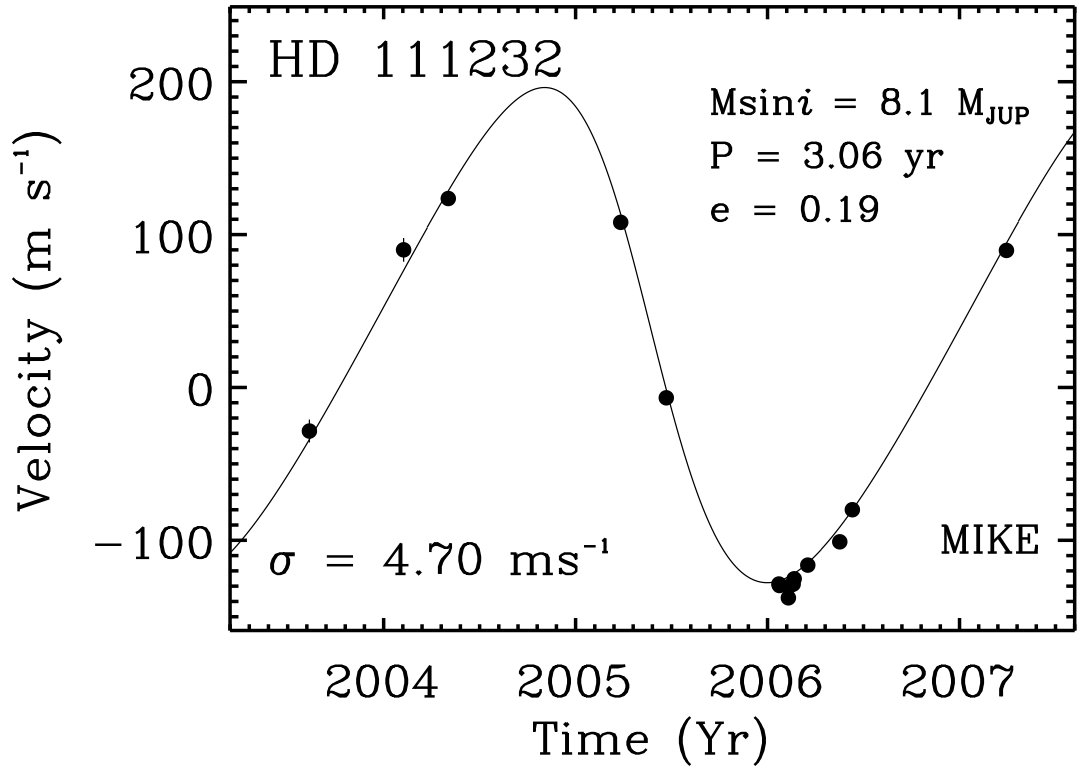}
\caption{ 
Doppler velocities for HD 111232 (G5 V).
The solid line is a Keplerian orbital fit with a
period of 1118 days, a semi-amplitude of 172 \ms,
and an eccentricity of 0.19, yielding a minimum
(\msini) of 8.1 \mjup \ for the companion.  The
RMS of the Keplerian fit is 4.70 \ms.}
\label{fig2}
\end{figure}

\begin{figure}
\includegraphics[angle=90,width=\textwidth]{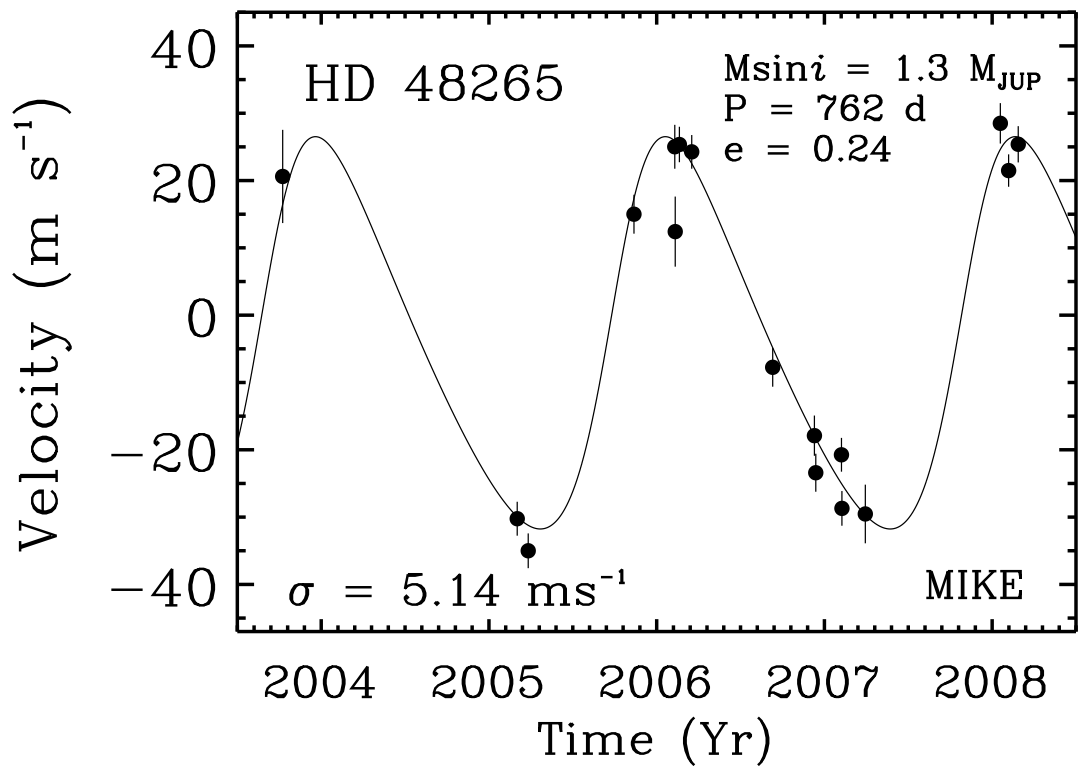}
\caption{Doppler velocities for HD 48265 (G5 V).
The solid line is a Keplerian orbital fit with a
period of 762 days, a semi-amplitude of 29 \ms,
and an eccentricity of 0.24, yielding a minimum
(\msini) of 1.3 \mjup \ for the companion.  The
RMS of the Keplerian fit is 5.14 \ms.}
\label{fig3}
\end{figure}

\begin{figure}
\includegraphics[angle=90,width=\textwidth]{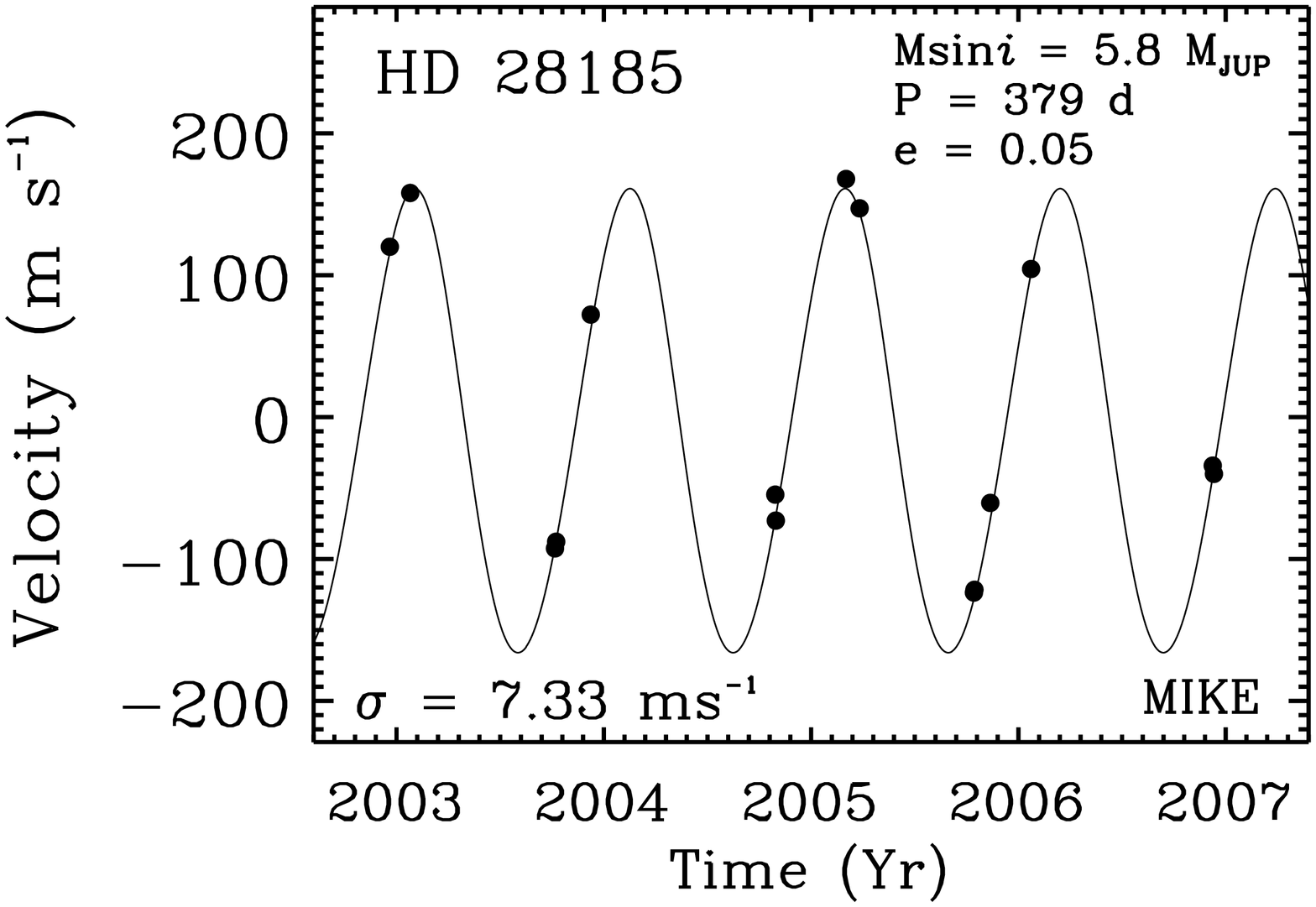}
\caption{Doppler velocities for HD 28185 (G0 V).
The solid line is a Keplerian orbital fit with a
period of 379 days, a semi-amplitude of 163 \ms,
and an eccentricity of 0.05, yielding a minimum
(\msini) of 5.8 \mjup \ for the companion.  The
RMS of the Keplerian fit is 7.33 \ms.}
\label{fig4}
\end{figure}

\begin{figure}
\includegraphics[angle=90,width=\textwidth]{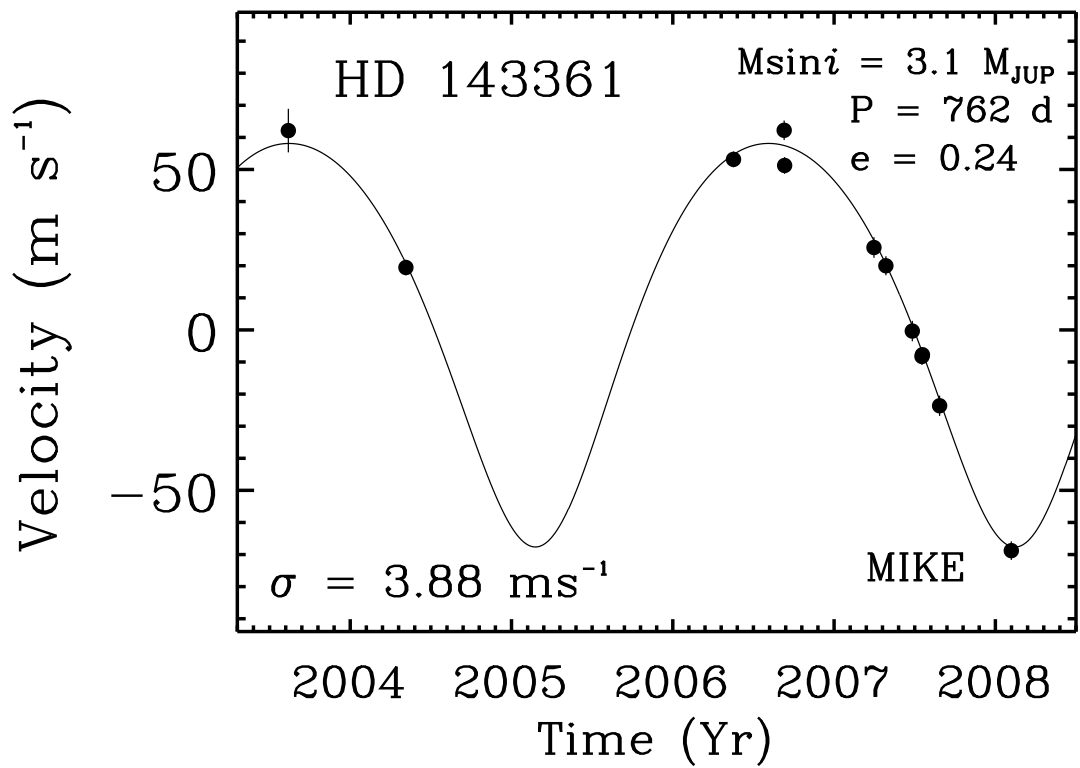}
\caption{Doppler velocities for HD 143361 (G0 V).
The solid line is a Keplerian orbital fit with a
period of 1086 days, a semi-amplitude of 63 \ms,
and an eccentricity of 0.18, yielding a minimum
(\msini) of 3.1 \mjup \ for the companion.  The
RMS of the Keplerian fit is 3.88 \ms.}
\label{fig5}
\end{figure}

\begin{figure}
\includegraphics[angle=90,width=\textwidth]{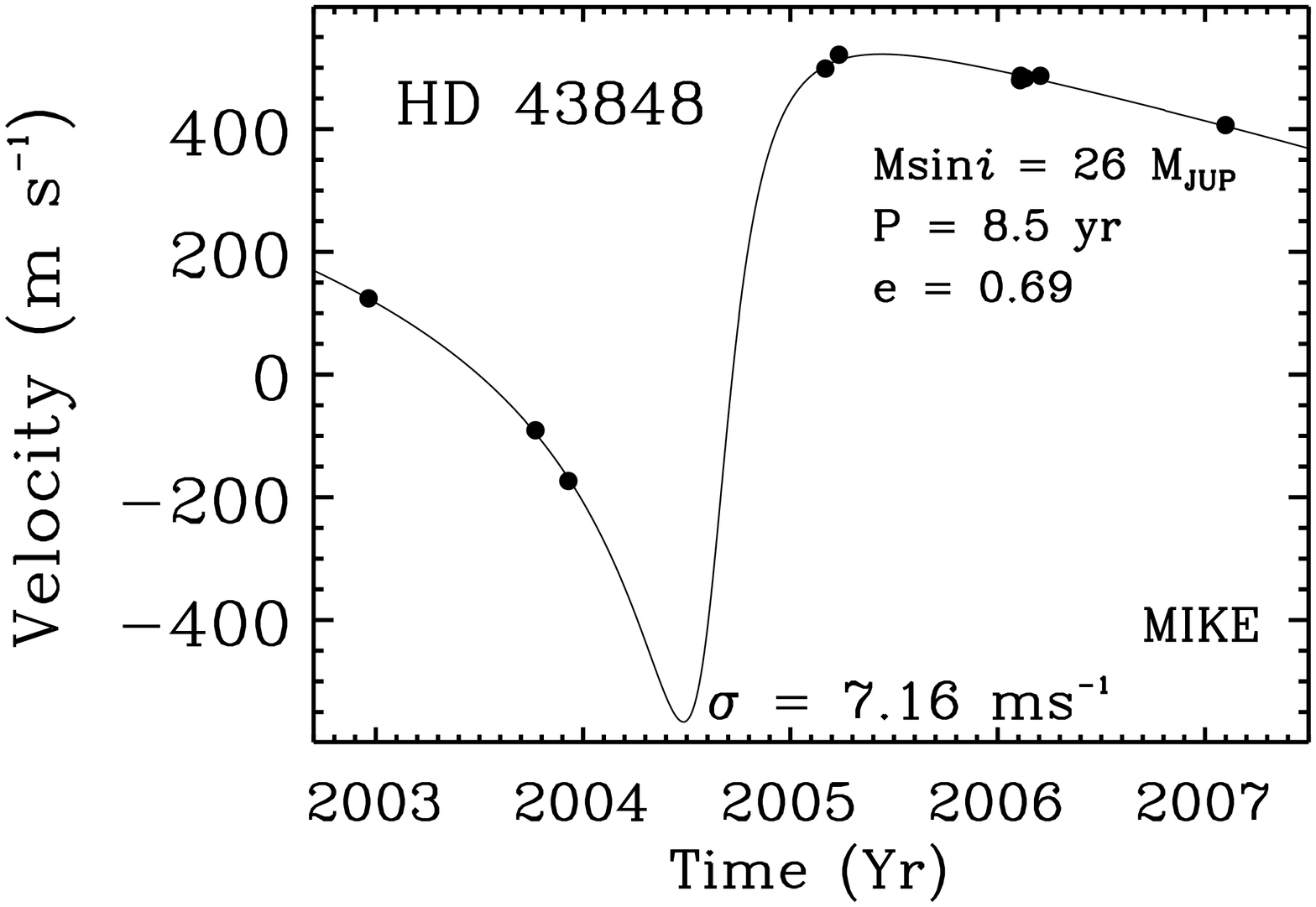}
\caption{Doppler velocities for HD 43848 (G2 V).
The solid line is a Keplerian orbital fit with a
period of 2371 days, a semi-amplitude of 544 \ms,
and an eccentricity of 0.69, yielding a minimum
(\msini) of 26 \mjup \ for the companion.  The
RMS of the Keplerian fit is 7.16 \ms.}
\label{fig6_01}
\end{figure}

\begin{figure}
\plotone{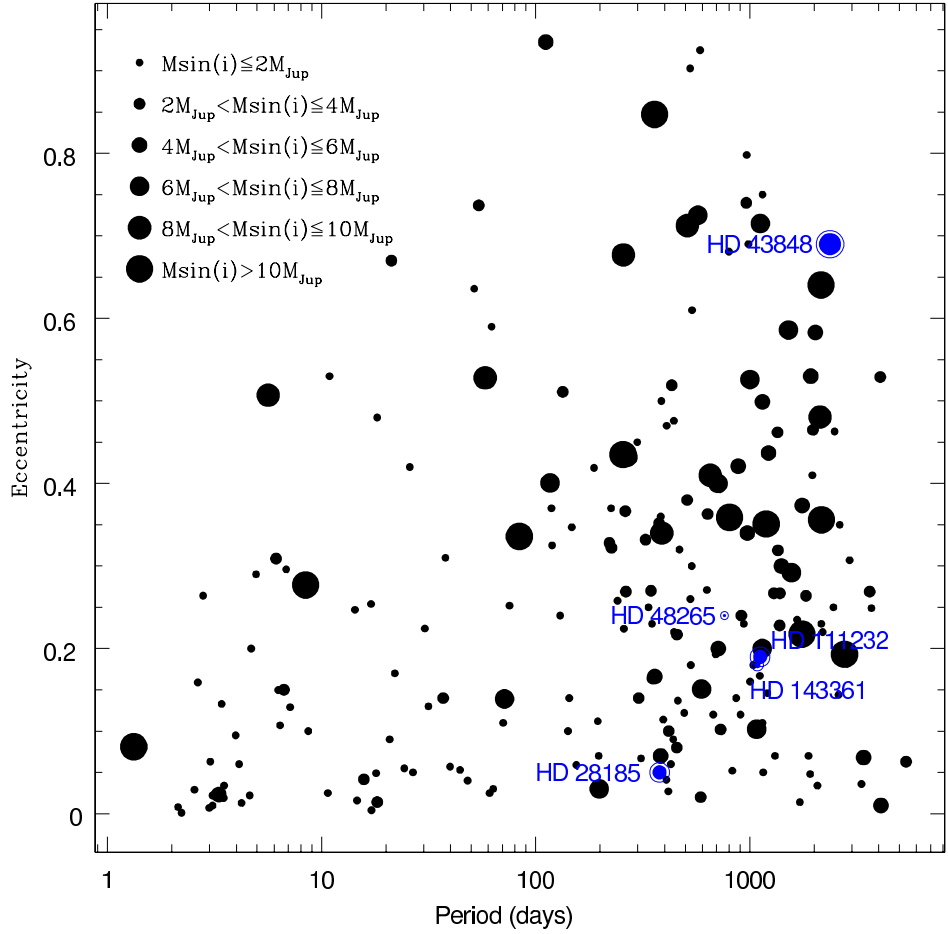}
\caption{Orbital eccentricity  $versus$ period for known extrasolar
planetary hosts. The circle sizes are proportional to planet mass 
as indicated.  The objects studied here are shown with large full
circles. Note the extreme
position of HD 43848, as a very massive and eccentric object.
}
\label{fig6_02}
\end{figure}

\begin{figure}
\plotone{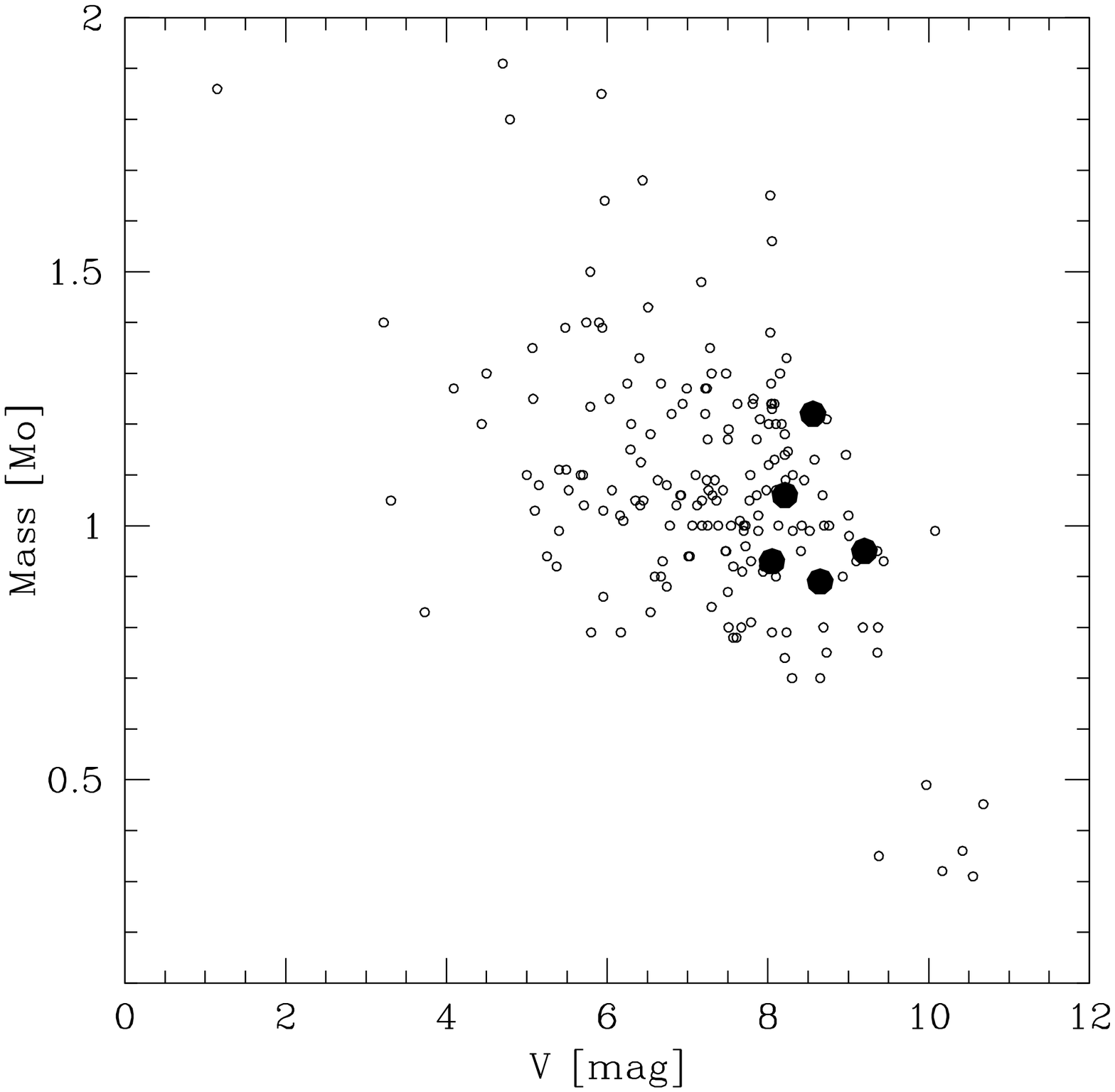}
\caption{Star mass $versus$ apparent $V$ magnitude for known extrasolar
planetary hosts (small circles). The new planets from our survey
(three from this work plus two from Lopez-Moralez et al. 2008) are 
shown with large full
circles. These first Magellan discoveries are among the faintest targets
surveyed for planets, probing deeper in the Solar Neighborhood.
The two previous CORALIE detections are included among the small circles.
}
\label{fig6_03}
\end{figure}


\begin{thebibliography}{}
\parsep 0pt
\itemsep -3pt

\bibitem[Baliunas { et~al.} 1995]{Bal95}
Baliunas, S.~L., Donahue, R.~A., Soon, W.~H., Horne, J.~H.,
Frazer, J., Woodard-Eklund, L., Bradford, M., Rao, L.~M.,
Wilson, O.~C., Zhang, Q., Bennett, W., Briggs, J.,
Carroll, S.~M., Duncan, D.~K., Figueroa, D., Lanning, H.~H.,
Misch, A., Mueller, J., Noyes, R.~W., Poppe, D., Porter, A.~C.,
Robinson, C.~R., Russell, J., Shelton, J.~C., Soyumer, T.
Vaughan, A.~H., \& Whitney, J.~H. 1995,
\newblock {  ApJ, } {438}, 269.

\bibitem[Bernstein { et~al.} 2003]{Ber03} Bernstein, R., et al. 2003,
\newblock { SPIE, } {4841}, 1694

\bibitem[Butler { et~al.} 1996]{BuMaWi96} Butler, R.~P., Marcy, G.~W.,
Williams, E., McCarthy, C., Dosanjh, P., \& Vogt, S.~S. 1996,
\newblock { PASP, } {108}, 500

\bibitem[Butler {  et~al.} 1999]{BuMaFi99}
Butler, R.~P., Marcy, G.~W., Fischer, D.~A., Brown, T.~M.,
Contos, A.~R., Korzennik, S.~G., Nisenson, P. \& Noyes, R.~W. 1999,
\newblock {  ApJ, } {526}, 916.

\bibitem[Butler { et~al.} 2006]{BuMa06}
Butler, R.~P. and Wright, J.~T. and Marcy, G.~W. and Fischer, D.~A.,  
	Vogt, S.~S. and Tinney, C.~G., Jones, H.~R.~A. 
	Carter, B.~D., Johnson, J.~A., McCarthy, C. \& 
	Penny, A.~J., 2006,
\newblock { ApJ, } {646}, 505.

\bibitem[Charbonneau { et~al.} 2000]{Char00}
Charbonneau, D., Brown, T.~M., Latham, D.~W. \& Mayor, M. 2000,
\newblock { ApJ, } {529}, L49.

\bibitem[Crane { et~al.} 2006]{Cra06} Crane, J., Shectman, S. A., \& Butler, R. P. 2006,
\newblock { SPIE, } {6269}, 96

\bibitem[Crane { et~al.} 2008]{Cra08} Crane, J., et al. 2008,
\newblock { SPIE, } in press

\bibitem[Eggenberger { et~al.} 2007]{Egg07} Eggenberger, A., et al. 2007,
\newblock { A\&A, } {474}, 273

\bibitem[ESA 1997]{ESA97}
ESA 1997, The Hipparcos and Tycho Catalogues (ESA SP-1200).

\bibitem[Fischer {  et~al.} 2005]{Fis05}
Fischer, D.~A., et al. 2003,
\newblock {  ApJ, } 620, 481

\bibitem[Fischer {  et~al.} 2005]{Fisch05}
Fischer, D.~A., et al. 2005,
\newblock {  ApJ, } 622, 1102

\bibitem[Fuhrmann 1998]{Fuhr98}
Fuhrmann, K. 1998, {  A\&A}, 338, 161.

\bibitem[Fuhrmann {  et~al.} 1997]{Fuhr97}
Fuhrmann, K., Pfeiffer, M.J. \& Bernkopf, J. 1997,
\newblock {  A\&A, } {326}, 1081.

\bibitem[Henry {  et~al.} 2000]{Henry00}
Henry, G.~W., Marcy, G.~W., Butler, R.~P. \& Vogt, S.~S. 2000,
\newblock { ApJ, } {529}, L45.

\bibitem[Lopez-Morales { et~al.} 2008]{LM08} Lopez-Morales, M., et al. 2008,
\newblock { AJ,} submitted

\bibitem[Marcy \& Butler 1992]{MaBu92}
Marcy, G.~W. \&  Butler, R.~P. 1992,
\newblock {  PASP, } {104}, 270.

\bibitem[Mayor \& Queloz 1995]{Mayor95}
Mayor, M., \& Queloz, D. 1995, Nature, 378, 355

\bibitem[Mayor { et~al.} 2004]{May04} Mayor, M., et al. 2004,
\newblock { A\&A,} {415}, 391

\bibitem[Nordstrom et al. 2004]{Nord04}
Nordstrom, B., et al. 2004, { A\&A,} 418, 989.

\bibitem[Perryman et al. 1996]{Perry96}
Perryman, M.~A.~C., et al. 1996, { A\&A,} 310, L21. 

\bibitem[Perryman et al. 1997]{Perry97}
Perryman, M.~A.~C., et al. 1997, { A\&A,} 323, L49. The Hipparcos Catalog

\bibitem[Pourbaix & Arenou 2001]{Pour2001}
Pourbaix, D., \& Arenou, F. 2001,
\newblock {  A\&A,} 372, 935.

\bibitem[Prieto & Lambert 1999]{Priet99}
Prieto, C.~A. \& Lambert, D.~L. 1999, { A\&A,} 352, 555.

\bibitem[Santos {  et~al.} 2001]{Santos01}
Santos, N. C., et al.  2001, A\&A, 379, 999

\bibitem[Valenti { et~al.} 1995]{VaBuMa95} 
Valenti, J., Butler, R.~P. \& Marcy, G.~W. 1995,
\newblock { PASP, } {107}, 966.

\end{thebibliography}
\end{document}